\pdfoutput=1
\documentclass[11pt]{article}
\usepackage{amsmath}
\usepackage{amssymb}
\usepackage{geometry}
\usepackage{tensor}
\usepackage{hyperref}
\usepackage{titlesec}
\usepackage{titling}
\usepackage{color}
\usepackage[title,titletoc]{appendix}
\usepackage[style=phys,biblabel=brackets,eprint=true]{biblatex}
\newcommand{\eqnref}[1]{Eq.~(\ref{#1})}
\title{Holographic models of non-Fermi liquid metals revisited: an effective field theory approach}
\begin{document}
\author{Dominic V. Else \\ \textit{Perimeter Institute for Theoretical Physics}}
\begin{titlingpage}
\maketitle
\begin{abstract}
    Accessing the physics of strongly coupled metals in a controlled way is a challenging problem in theoretical condensed matter physics. In this paper, we revisit the possibility of understanding strongly coupled metals through a holographic duality with a weakly coupled gravitational theory in one higher dimension (i.e.\ a suitable generalization of the ``AdS/CFT duality''). Previous attempts at devising holographic models of strongly coupled metals have suffered from severe drawbacks; for example, they do not even seem to be able to describe a Fermi surface that satisfies Luttinger's theorem, which is ought to be a core requirement in any physically reasonable model of a metal. Here, we propose a radically different approach to constructing holographic models of strongly coupled metals. The idea is that for applications, it should be sufficient to construct a holographic dual of the effective field theory that controls the infra-red physics of the metal. We invoke recent work that has identified a precise criterion for such an effective field theory to be ``emergeable'' from a continuum ultra-violet (UV) theory at nonzero charge density (or its equivalent in lattice models, namely an incommensurate charge filling). We show that imposing this criterion leads to a holographic model of a strongly coupled metal with physically reasonable properties, including a Fermi surface satisfying Luttinger's theorem. We discuss a possible physical interpretation of our results.
\end{abstract}
\end{titlingpage}
\tableofcontents
\newpage
\section{Introduction}
Understanding strongly coupled quantum many-body phases of matter is a crucial problem in condensed matter physics. One important class of such phases of matter are the so-called ``non-Fermi liquids'', which are metals that are not described by the conventional weakly-coupled Fermi liquid theory. Non-Fermi liquid physics is believed, for example, to be behind the exotic ``strange-metal'' regime seen in high-$T_c$ cuprates \cite{Proust_1807,Varma_1908} as well as other classes of materials \cite{Lohneysen_0606,Gegenwart_0712,Cao_1901}.

The strongly-coupled nature of non-Fermi liquids has made it challenging to find models in which any physics can be obtained in a controlled way.
One seemingly appealing strategy would be to invoke the idea of holography, or ``AdS/CFT'' \cite{HQM}, in which certain strongly coupled quantum field theories (QFTs) are held to be dual to a \emph{weakly} coupled quantum gravity theory in one higher space-time dimension. In an appropriate limit of the QFT, the dual gravitational theory can be treated classically, and the physics of the strongly coupled QFT can be extracted simply by solving the classical equations of motion in the dual theory.
Highly non-trivial quantum many-body effects in the QFT, such as thermalization and dissipation, can be ``geometrized'', originating in the dual theory from the presence of a black hole.

Although such an approach has led to powerful insights in other areas \cite{Kovtun_0405,Maldacena_1503,Kim_1107}, the situation for non-Fermi liquid metals is not very satisfactory, despite a plethora of studies. The usual approach is that one starts from some gravitational theory that is supposed to be dual to a strongly coupled conformal field theory (CFT) (or a more general strongly coupled gapless theory without Lorentz or conformal invariance), and then imagines perturbing the field theory by switching on a nonzero charge density, leading to an RG flow to some new infra-red (IR) fixed-point which presumably describes some kind of strongly coupled metal. In the dual theory this corresponds to introducing an electric field, which backreacts on the metric, inducing a new geometry.

Unfortunately such models seem to inevitably have various pathologies. In the simplest model \cite{Chamblin_9902,Lee_0809,Liu_0903,Cubrovic_0904,Faulkner_0907}, the bulk gravitational geometry is the so-called AdS-Reissner-Nordstr\"om metric, and the IR regime contains a charged black hole. The problem is that the Bekenstein-Hawking entropy of the black hole implies that the dual QFT has a nonzero entropy density even at zero temperature. Although this is of course similar to what happens in the Sachdev-Ye-Kitaev (SYK) model \cite{Georges_0009,Fu_1603}, it seemingly contradicts the Third Law of Thermodynamics and seems very unlikely to occur in a realistic system without fine-tuning. By considering variants of this model with different values of the dynamical critical exponent and hyperscaling violation exponent \cite{Goldstein_0911,Gubser_0911,Huijse_1112,HQM} it is possible to eliminate the zero temperature entropy density, but this often comes at the expense of introducing other pathologies such as naked singularities in the gravitational theory (although these singularities may be considered acceptable \cite{Charmousis_1005} in the sense that they could be resolvable in a quantum gravity theory).

From our point of view, however, the most serious issue with these models is that they do not seem to capture the Fermi surface. In Fermi liquid theory, the ``Fermi surface'' -- the codimension-1 surface in momentum space where the low-energy quasiparticles live -- is crucial to the physics. Although non-Fermi liquids generally do not have quasiparticles, to the extent that we understand non-Fermi liquid physics in non-holographic models (for example, the ``Hertz-Millis'' type theories of quantum critical points \cite{Hertz__1976,Millis__1993,Lee_1703}), a generalized notion of Fermi surface still appears to be key to the physics. Another important aspect of the Fermi surface is Luttinger's theorem \cite{Luttinger__1960,Oshikawa_0002}, which relates the volume enclosed by the Fermi surface to the microscopic charge density; in a rough sense, one should think of the portion of momentum space enclosed by the Fermi surface, known as the ``Fermi sea'', as being ``where the UV charge goes in the IR''. Although originally described for Fermi liquid theory, Luttinger's theorem is now understood to be much more general \cite{Else_2007}.

Although in some cases one finds Fermi surfaces in holographic models of non-Fermi liquids \cite{Lee_0809,Liu_0903,Cubrovic_0904,Faulkner_0907}, they are generally ``small'' Fermi surfaces that do not satisfy Luttinger's theorem on their own, raising the question of what happened to the remainder of the UV charge (this can be traced back to the fact that most of the charge in the gravitational theory is hidden inside the black hole \cite{Iqbal_1112}). Attempts have been made \cite{Huijse_1104,Huijse_1112,Iqbal_1112} to draw analogies with phenomena known in condensed matter, such as (a) ``fractionalized'' phases such as the so-called fractionalized Fermi liquid (``FL*'') \cite{Senthil_0209,Senthil_0305} in which a portion of the charge density is attributed to the topological degrees of freedom and does not contribute to the Fermi surface volume; or (b) systems such as the ``composite Fermi liquid'' \cite{Halperin__1993,Son_1502} in which the Fermi surface relates to particles which are charged under an emergent deconfined gauge field, and hence is ``hidden'', i.e.\ not easily detectable by gauge-invariant operators.
 However, we feel that such analogies are highly misleading. In both cases (a) and (b), the system has only a microscopic lattice translation symmetry\footnote{In some cases, such as the composite Fermi liquid, the symmetry can be extended to a continuous translation symmetry, but one which is ``magnetic'', i.e.\ the symmetry group is a non-trivial central extension of $\mathbb{R}^d$ by $\mathrm{U}(1)$, as occurs in a uniform magnetic field. We do not want to consider such magnetic translation symmetries. The lattice translation symmetry we are referring would correspond to a commuting subgroup of the full non-commutative magnetic translation symmetry.}. One can show using the general methods of Refs.~\cite{Paramekanti_0406,Bonderson_1601,Else_2007}  that these mechanisms for a violation of Luttinger's theorem \emph{cannot} be extended to a system with microscopic continuous translation symmetry. Therefore, in systems with microscopic translation symmetry, that are not superconducting, Luttinger's theorem should be considered a non-negotiable requirement, in contradiction to what one seems to find in the holographic models. Therefore, the fact that these holographic models cannot, for example, reproduce all the transport propreties of experimental ``strange metals'' \cite{Amoretti_1603,Ahn_2307} should not be surprising, as they are not even building in the most basic aspects of the physics.

There are some suggestions that one does recover a Fermi surface satisfying Luttinger's theorem if one considers quantum gravity corrections in the gravitational theory \cite{Polchinski_1203, Faulkner_1207}. On the other hand, since the Fermi surface is presumably central to the low-energy physics, this eliminates much of the original appeal of the holographic approach, namely that one can understand the physics of a strongly coupled system solely by solving classical equations of motion.

\subsection{A new approach: holographic effective field theory}

In this work, we wish to advocate an alternative approach to developing holographic models of non-Fermi liquids. In condensed matter physics, one normally does not try to exactly solve a microscopic lattice model at all scales. Instead, one invokes the concept of emergence -- the IR physics, i.e.\ the physics at sufficiently long wavelengths, low frequency, and low temperature should be captured by an effective field theory, and one seeks to understand the nature of this effective field theory and not to worry about how exactly it emerges from the microscopic model. In the language of RG, the microscopic lattice model can be viewed as a UV theory which flows to a stable fixed-point in the IR, and one seeks to understand this IR fixed-point, not the details of how exactly the RG flow runs starting from the UV. indeed, Fermi liquid theory itself is best viewed from this perspective \cite{Polchinski_9210,Shankar_9307}.

In holography, the additional spatial coordinate in the higher-dimensional space-time can be interpreted with respect to the dual QFT as an ``RG parameter''. Thus, in the holographic models discussed previously, what one is effectively attempting to do is to take a UV theory (e.g. some strongly coupled CFT), perturb it in some way (by switching on a nonzero charge density) and then study the entire RG flow from the UV theory (corresponding to near-boundary region  of the bulk space-time) to the IR fixed point (corresponding to the region of the bulk space-time far away from the boundary). This is much more ambitious than what one typically attempts to do in condensed matter physics. Moreover, the relevance to condensed matter physics is in any case limited, since in condensed matter the UV theory will always be some lattice model, not a continuum field theory.

Therefore, what we advocate in this paper is to give up on this goal, and instead come up with a holographic formulation of a plausible \emph{IR effective field theory} of a metal. This raises the obvious question, however, of what criteria we should use to judge a potential IR theory. Ultimately, of course, one must judge it by comparisons to experiment. However, in the meantime a useful criterion is the one which has been dubbed ``emergeability'' \cite{Zou_2101}: given a lattice model with certain properties (e.g.\ symmetries such as charge conservation and lattice translation symmetry), under which circumstances is it theoretically possible for a given effective field theory to arise as the low-energy description of the lattice model? Specifically, there are certain matching conditions between the UV and IR that must be satisfied.

An important example of such matching conditions are the so-called ``filling constraints'' \cite{Oshikawa_9610,Oshikawa_9911,Hastings_0411,Luttinger__1960,Oshikawa_0002,Paramekanti_0406,Bonderson_1601,Else_2007,Else_2106}. If we have a lattice system in $d$ spatial dimensions with $\mathrm{U}(1)$ charge conservation symmetry and $\mathbb{Z}^d$ lattice translation symmetry, then one can define a real number $\nu$, called the \emph{filling} which describes the average charge per unit cell in the ground state. In general there is a matching condition between the fractional part of $\nu$ and properties of the low-energy theory. An example of such a constraint is Luttinger's theorem, which we already mentioned above; in the case of lattice translation symmetry, the precise statement is that in a spinless Fermi liquid,
\begin{equation}
    \frac{\mathcal{V}_F V_{\mathrm{unit}}}{(2\pi)^d} = \nu \quad \mathrm{mod} \, 1,
\end{equation}
where $\mathcal{V}_F$ is the volume in momentum space enclosed by the Fermi surface, and $V_{\mathrm{unit}}$ is the volume of a translation unit cell.

A particularly interesting case is when the filling $\nu$ can be tuned to be an \emph{irrational} number; we call such systems ``compressible''. Compressibility implies very strong constraint on the low-energy physics \cite{Else_2007,Else_2010,Else_2106}. Specifically, it was argued in Refs.~\cite{Else_2007,Else_2106} that the only way for the IR theory to be compatible with compressibility in spatial dimension $d > 1$ is that either there must be an emergent higher-form symmetry, or there must be an infinite-dimensional emergent symmetry group. The former possibility is realized in superfluids where the charge $\mathrm{U}(1)$ is spontaneously broken and there is an emergent $(d-1)$-form symmetry. The latter possibility is realized in Fermi liquid theory, where in the IR theory the charge at \emph{every} point on the Fermi surface is separately conserved, corresponding to an infinite-dimensional symmetry group.

An empirical observation that one can make is that all metals, including non-Fermi liquids seem to be compressible. Therefore, in seeking to identify a plausible IR theory for a metal, it is reasonable to demand that it should be compatible with compressibility. In particular, we can consider systems in which the compressibility is activated in the same way as in Fermi liquid theory, through an infinite-dimensional symmetry group (which for simplicity, we will assume takes the same form as in Fermi liquid theory). Such IR theories were referred to in Ref.~\cite{Else_2007} as ``ersatz Fermi liquids''. Thus, we arrive at the main goal of this paper: to formulate a holographic model of an ersatz Fermi liquid.

What we will see is that such an approach indeed allows us to obtain a holographic model that seems to have physically reasonable properties for a metal, more so than previous holographic models. Moreover, a key advantage of our model is that unlike previous holographic models, it explicitly builds in a Fermi surface (that satisfies Luttinger's theorem).

\subsection{Outline}
The remainder of the paper is organized as follows. In Section \ref{sec:efl_review}, we review general properties of ersatz Fermi liquids. In Section \ref{sec:model}, we define the holographic model of an ersatz Fermi liquid we are considering. In Section \ref{sec:results}, we present the results from a solution of the model. In Section \ref{sec:interpretation} we discuss a possible intepretation of the model in terms of a characterization of the dual QFT. In Section \ref{sec:entanglement}, we discuss the scaling of entanglement entropy and charge fluctuations in the ground state and compare with Fermi liquid theory. Finally, in Section \ref{sec:outlook} we discuss future directions.

\section{Review: ersatz Fermi liquids}
\label{sec:efl_review}

\subsection{Emergent symmetry, conservation laws, and 't Hooft anomaly}
\label{subsec:emergent_symmetry}
Fermi liquids, and hence, by definition, ersatz Fermi liquids, have an infinite-dimensional emergent symmetry group, which, in $d=2$ spatial dimensions, we call $\mathrm{LU}(1)$ \cite{Else_2007}. It is an example of what mathematicians call a ``loop group''. Specifically, $\mathrm{LU}(1)$ is the group comprising all smooth functions from the circle $S^1$ into $\mathrm{U}(1)$. [The group law applies pointwise, i.e.\ if $f,g \in \mathrm{LU}(1)$ are functions from $S^1$ into $\mathrm{U}(1)$, then $(f \cdot g)(s) = f(s) g(s)$, where the right-hand side refers to multiplication in $\mathrm{U}(1)$]. Roughly, the fact that $\mathrm{LU}(1)$ is an emergent symmetry reflects the fact that the charge at every point on the Fermi surface is individually conserved -- in Fermi liquid theory this is attributed to the absence of quasiparticle scattering (that is, the interactions that would lead to such scattering are irrelevant in the RG sense). The circle $S^1$ represents the Fermi surface. In this paper we will parameterize the circle, and hence the Fermi surface, by a coordinate $\theta$ (all of the statements we make will hold for an arbitrary parameterization). Notice that $\mathrm{LU}(1)$ contains a $\mathrm{U}(1)$ subgroup comprising the constant functions; we can identify this with the microscopic $\mathrm{U}(1)$ charge conservation symmetry.

The charges of $\mathrm{LU}(1)$ correspond to irreducible representations, which, since the group is Abelian, are 1-dimensional. Such irreps can be labelled by real-valued distributions\footnote{That is to say, real-valued functions, except that we also allow proper distributions such as delta functions.} $N(\theta)$, such that an element $f \in \mathrm{LU}(1)$ acts as a phase factor
\begin{equation}
    \label{eq:irrep}
    \exp \left( i \int f(\theta) N(\theta) d\theta \right),
\end{equation}
where here we view the $\mathrm{U}(1)$ target of elements of $\mathrm{LU}(1)$ as $\mathbb{R}/(2\pi\mathbb{Z})$. The fact that $f(\theta)$ has a mod $2\pi$ ambiguity requires us to impose the condition that $\int N(\theta) d\theta$ is an integer to ensure that the phase factor \eqnref{eq:irrep} is well-defined. Physically, $N(\theta)$ can be interpreted as the charge distribution on the Fermi surface, such that $\int N(\theta) d\theta$ is the total $\mathrm{U}(1)$ charge. Going beyond the 1-dimensional irreps, we can define an operator-valued distribution $\hat{N}(\theta)$ such that an element $f \in \mathrm{LU}(1)$ acts on the whole Hilbert space as
\begin{equation}
    \exp \left( i \int f(\theta) \hat{N}(\theta) d\theta \right).
\end{equation}
We can (roughly) think of $\hat{N}(\theta)$ as the generators of the action of  $\mathrm{LU}(1)$ on the Hilbert space, and viewed as observables they measure the (conserved) charge distribution on the Fermi surface. In the rest of the paper we will drop the hats on $\hat{N}(\theta)$.

In Fermi liquid theory the emergent $\mathrm{LU}(1)$ symmetry has a so-called 't Hooft anomaly, meaning that there is an obstruction to gauging the symmetry. This is reflected in the fact that when a background gauge field of the $\mathrm{LU}(1)$ symmetry is applied, the $\mathrm{LU}(1)$ charge can become non-conserved. In order to explain this, 
let us first define what we mean by an $\mathrm{LU}(1)$ gauge field. In general, a gauge field for a continuous group on a space-time $M$ is a covariant vector field on $M$ valued in the algebra of infinitesimal transformations of the group. Concretely, given the definition of $\mathrm{LU}(1)$, this suggests that an $\mathrm{LU}(1)$ gauge field on $M$ is a family $A_\mu(\theta)$ of covariant vector fields on $M$ that smoothly depends on the parameter $\theta \in S^1$, with the gauge transformation
\begin{equation}
    A_\mu(\theta) \to A_\mu(\theta) + \partial_\mu \lambda(\theta),
\end{equation}

In fact, however, as pointed out in Ref.~\cite{Else_2007}, this is not the entire story -- there is an additional wrinkle in the definition of gauge field that applies only to infinite-dimensional groups such as $\mathrm{LU}(1)$. One actually needs to include an additional component $A_\theta$ that transforms under gauge transformations as $A_\theta \to A_\theta + \partial_\theta \lambda$. In Fermi liquid theory, where one can talk about quasiparticles that are localized both in space and in momentum space, the spatial components of $A$ describe the quantum phase accumulated as the quasiparticle is moved in space, while $A_\theta$ describes the quantum phase accumulated as the quasiparticle is moved along the Fermi surface in momentum space. [One can argue that $A_\theta$ is still a necessary ingredient for an $\mathrm{LU}(1)$ gauge field even beyond Fermi liquid theory.]
We can now make the observation that an $\mathrm{LU}(1)$ gauge field on $M$ looks formally equivalent to a $\mathrm{U}(1)$ gauge field on a higher-dimensional space $M \times S^1$ (one should be careful, however, about taking this analogy too far, as we will see later).

We can now state the nature of a 't Hooft anomaly of the $\mathrm{LU}(1)$ symmetry \cite{Else_2007}. We can introduce the $\mathrm{LU}(1)$ current $j^\mu$, which is a contravariant vector field on $M \times S^1$. For example, one could define $j^\mu = \frac{\delta S}{\delta A_\mu}$, where $S$ is the action of the system coupled to the $\mathrm{LU}(1)$ gauge field [in particular, in principle $j$ includes a component $j^\theta$; we discuss this further below.] The time component $j^t$ can be viewed as the spatial density of the $N(\theta)$ defined above.
 Then the anomaly equation takes the form
\begin{equation}
    \label{eq:conservation_eqn}
    \partial_\mu j^\mu = \frac{m}{8\pi^2} \epsilon^{\mu \nu \lambda \sigma} [\partial_\mu A_\nu] [\partial_\lambda A_\sigma].
\end{equation}
Note that in these equations, we allow the greek-letter indices to vary not just over the directions of space-time, but also over the $\theta$ coordinate (hence how we are able to use the 4-dimensional Levi-Civita symbol $\epsilon$, even though we began with a 3-dimensional space-time). The anomaly coefficient $m$ is quantized to be an integer through general arguments; in single-component Fermi liquid theory it takes the values $\pm 1$ depending on an (arbitrary) choice of orientation of the Fermi surface. Observe that this anomaly equation has the same structure as for a $\mathrm{U}(1)$ gauge field in a 4-dimensional space-time; for a $\mathrm{LU}(1)$ gauge field in 3-dimensional space-time, the Fermi surface plays the role of an ``extra dimension''.

Finally, let us return to the issue of the $j^\theta$ component of the current. In order to really be able to say that the system has an $\mathrm{LU}(1)$ symmetry, $j^\theta$ must obey some strong restrictions. If $j^\theta$ is nonzero it implies a flow of charge along the Fermi surface. In general this will imply that the total charge $N(\theta)$ at each point on the Fermi surface is no longer conserved individually. Therefore, a system with $\mathrm{LU}(1)$ symmetry must obey the property that $j^\theta$ is identically zero. An exception to this could occur in the presence of a magnetic field; for example it is well known that in Fermi liquid theory, a magnetic field induces a precession of quasiparticles along the Fermi surface, which would correspond to $j^\theta \neq 0$. This allows for the $N(\theta)$ to become non-conserved in the presence of a magnetic field. This may not be too shocking given the 't Hooft anomaly, but we note that in this case the non-conservation actually arises from the $\partial_\theta j^\theta$ term in \eqnref{eq:conservation_eqn}, not the right-hand side of \eqnref{eq:conservation_eqn} as one might have expected.

\subsection{Fermi surface and phase space magnetic field}
\label{subsec:phase_space_magnetic_field}
We can write the anomaly equation \eqnref{eq:conservation_eqn} as
\begin{equation}
    \label{eq:expanded_anomaly_eqn}
    \partial_\mu j^\mu = \frac{m}{(2\pi)^2} [ B F_{\theta t} + \epsilon^{ij} E_i F_{\theta j}],
\end{equation}
where we defined the field strength tensor $F_{\mu \nu} = \partial_\mu A_\nu - \partial_\nu A_\mu$; $t$ denotes the time direction; $i$ and $j$ range over the two spatial directions; and we have defined the magnetic field $B = \frac{1}{2} \epsilon^{ij} F_{ij}$ and electric field $E_i = F_{ti}$.
If we set $E_i$ and $B$ to be independent of $\theta$, this will correspond to applying a background gauge field for the $\mathrm{U}(1)$ subgroup of $\mathrm{LU}(1)$. Suppose in particular that we just consider an electric field and set $B=0$. In that case, it is known that in Fermi liquid theory (in which we can set the anomaly coefficient $m=1$), the non-conservation of $\mathrm{LU}(1)$ charge takes the form
\begin{equation}
    \label{eq:fermi_liquid_anomaly}
    \partial_\mu j^\mu = \frac{1}{(2\pi)^2} \epsilon^{ij} E_i \partial_\theta k_j(\theta),
\end{equation}
where the vector $\mathbf{k}(\theta)$ denotes the (vector) Fermi momentum as a function of position on the Fermi surface.

In order for \eqnref{eq:expanded_anomaly_eqn} and \eqnref{eq:fermi_liquid_anomaly} to agree, it appears that we must identify
\begin{equation}
    \label{eq:phase_space_magnetic_field}
F_{\theta j} = \partial_\theta k_j(\theta).
\end{equation}
The necessity of this identification was previously pointed out in Ref.~\cite{Else_2106} (and was somewhat implicit in Ref.~\cite{Else_2007}). A nice interpretation was suggested in Ref.~\cite{Lu_2302}: since moving in $\theta$ space amounts to moving along the Fermi surface, and the Fermi surface lives in momentum space, \eqnref{eq:phase_space_magnetic_field} reflects the non-commutativity between position and momentum coordinates, which can be encoded by a ``magnetic field'' in phase space.

We will take it for granted that the identification \eqnref{eq:phase_space_magnetic_field} will continue to hold even beyond Fermi liquid theory, in any ersatz Fermi liquid.
Indeed, in a general ersatz Fermi liquid we can simply \emph{define} Fermi surface in such a way that \eqnref{eq:phase_space_magnetic_field} is identically satisfied. More precisely, suppose we consider a translationally invariant configuration of the system; in that case, we should be able to choose a gauge such that $\partial_i A_\theta = 0$. Then \eqnref{eq:phase_space_magnetic_field} tells us that $\partial_\theta [A_i(\theta) - k_i(\theta)]= 0$, so we can define the Fermi surface momentum (up to an overall additive constant) according to $k_i(\theta) = A_i(\theta)$.

\subsection{Luttinger's theorem and compressibility}
\label{subsec:luttinger}
Suppose that our ersatz Fermi liquid, with emergent $\mathrm{LU}(1)$ symmetry, describes the emergent IR physics of a microscopic system that has a global $\mathrm{U}(1)$ symmetry, as well as either a lattice or continuous translation symmetry. Then it turns out that there is a ``UV-IR'' matching condition that one can derive between the properties of the IR theory and the microscopic density of the charge of the global $\mathrm{U}(1)$ symmetry \cite{Else_2007}. In the context of Fermi liquid theory, this is known as Luttinger's theorem. In the case of continuous translation symmetry, the relation takes the form
\begin{equation}
    \rho = \frac{m \mathcal{V}_F}{(2\pi)^2},
\end{equation}
where $\rho$ is the microscopic charge density, and $\mathcal{V}_F$ is the volume enclosed by the Fermi surface [recall from the previous subsection that in a general ersatz Fermi liquid, the Fermi surface can be defined in terms of the background $\mathrm{LU}(1)$ gauge field.] For a system with lattice translation symmetry, the statement instead takes the form
\begin{equation}
    \nu = \frac{m \mathcal{V}_F V_{\mathrm{unit}}} {(2\pi)^2} \quad [\operatorname{mod}  1],
\end{equation}
where $V_{\mathrm{unit}}$ is the volume of a translation unit cell, and the dimensionless number $\nu$, known as the ``filling'', is the average charge per unit cell.

From the above relations, we see that in the case of continuous microscopic translation symmetry, an ersatz Fermi liquid is compatible with a nonzero microscopic charge density; while in the case of discrete microscopic translation symmetry, an ersatz Fermi liquid is ``compressible'', in the sense that the microscopic filling $\nu$ can be continuously tuned simply by varying the Fermi surface volume.

\subsection{Hydrodynamics}
\label{subsec:hydrodynamic}
The infinitely many conservation laws of an ersatz Fermi liquid have very strong consequences for the dynamics. In particular, Ref,~\cite{Else_2301} studied the dynamics in the ``hydrodynamic'' regime. In this regime one assumes that the system is locally in thermal equilibrium at each point in space and time. Here the concept of ``thermal equilibrium'' needs to take into account all the conserved quantities. Thus, the local equilibrium state will depend on the local densities of the conserved quantities $N(\theta)$, which might vary as a function of space and time. Hydrodynamics gives an equation of motion for how these densities evolve in time.

Ref.~\cite{Else_2301} showed that, at zero-th order in a gradient expansion, and working to linear order in the perturbation from the global equilibrium state, one obtains, in a general ersatz Fermi liquid, an equation of motion that depends only on certain thermodynamic susceptibilities $\xi(\theta,\theta')$ of the conserved charges $N(\theta)$. Let us focus on the case where these susceptibilities contain only a contact term, i.e.\ $\xi(\theta,\theta') = v_F(\theta) \delta(\theta - \theta')$. This certainly need not be true in general (and is not even true in Fermi liquid theory when the Landau interactions are nonzero), but we will see later that it actually is what happens in our particular holographic model. In this case, the equations of motion of Ref.~\cite{Else_2301} reduce to
\begin{equation}
    \label{eq:hydrodynamic_eqn}
    \frac{\partial n(\theta)}{\partial t} + \mathbf{v}_F(\theta) \cdot \nabla n(\theta) = \frac{m}{(2\pi)^2}\mathbf{E} \cdot \mathbf{w}(\theta),
\end{equation}
where $\mathbf{E}$ is an applied background electric field, and we defined the vectors $\mathbf{w}(\theta)$ and $\mathbf{v}_F(\theta)$ according to $w^i(\theta) = \epsilon^{ij} \partial_\theta k_j(\theta)$ [recall that $\mathbf{k}(\theta)$ is the Fermi momentum vector], and $\mathbf{v}_F(\theta) = v_F(\theta) \mathbf{w}(\theta) / |\mathbf{w}(\theta)|$. This happens to be (if we set $m=1$) the same equations of motion that one would get in a Fermi liquid with the Landau interactions set to zero.

The fact that the derivation of Ref.~\cite{Else_2301} was based on hydrodynamics, which in turn is based on the assumption of local thermal equilibrium, suggests that there could in principle be some limitations to the validity of \eqnref{eq:hydrodynamic_eqn}. In particular, if we consider dynamics at frequency $\omega$, hydrodynamics does not necessarily apply when $\omega$ is larger than the inverse local thermalization time, for which $\sim T$ is a good guess at low temperatures in a strongly coupled system. Thus, in principle we should only expect \eqnref{eq:hydrodynamic_eqn} to hold when $\omega \ll T$. However, in Fermi liquid theory \eqnref{eq:hydrodynamic_eqn} actually holds without any such restriction; we will see that this also ends up being the case in our holographic model.

    \section{A holographic model of an ersatz Fermi liquid}
    \label{sec:model}
    \subsection{The bulk action}
    \label{subsec:bulk_action}
    We refer the reader to Ref.~\cite{HQM} for an accessible introduction to the basic framework of holographic models.
    In this paper, we wish to find a bulk gravitational theory that is holographically dual to a boundary QFT that has a global $\mathrm{LU}(1)$ symmetry.  According to the standard holographic dictionary, the way to achieve this is clear: we need the bulk theory to have a dynamical $\mathrm{LU}(1)$ gauge field.
    
    As mentioned in Section \ref{subsec:emergent_symmetry}, an $\mathrm{LU}(1)$ gauge field on a 4-dimensional space-time $M$ can in a certain sense be thought of as a vector field $A$ on the 5-dimensional space $M \times S^1$.
We emphasize, however, that the $S^1$ should \emph{not} be thought of as an additional, compactified dimension of space-time, such that, for example, the metric in the gravitational theory obeys the Einstein equations for the five-dimensional space-time. For one thing, if we formulated the holographic model in this way, it would imply that the boundary theory lives on the 4-dimensional space-time $\partial M \times S^1$. In particular, it would be possible to define a local energy density for the boundary theory on $\partial M \times S^1$. By contrast, in a metal with a Fermi surface, in general the Hamiltonian will couple different points on the Fermi surface without any regard to locality, so there is no such notion of a local energy density on $\partial M \times S^1$ -- only a local \emph{charge} density.
These considerations suggest that we should instead identify the 4-dimensional manifold $M$ as the space-time manifold, and require that the metric in the bulk gravitational theory obeys the Einstein equations on $M$.

A related subtlety is that we need to make sure that the gauge field in the bulk really can be interpreted as an $\mathrm{LU}(1)$ gauge field on $M$, rather than a $\mathrm{U}(1)$ gauge field on $M \times S^1$. According to the discussion at the end of Section \ref{subsec:emergent_symmetry}, this means that the action must have the property that $j^\theta = \frac{\delta S}{\delta A_\mu}$ is identically zero (at least in the absence of a magnetic field). If we just wrote down the Maxwell action for a $\mathrm{U}(1)$ gauge field on $M \times S^1$, it would not satisfy this property.

Instead, we will employ a Maxwell action of the form
\begin{align}
    \label{eq:maxwell}
    S_{\mathrm{Maxwell}} &= -\frac{1}{4} \int_{M \times S^1} \frac{1}{\alpha(\theta)} f_{\mu \nu} f^{\mu \nu} \sqrt{-g} d^4 x d\theta, \\
                          & \quad \quad \quad \quad \quad \quad f_{\mu \nu} = \partial_\mu a_\nu - \partial_\nu a_\mu, \nonumber
\end{align}
where here the Greek letters range over the 4 dimensions of the space-time manifold $M$, but not over the $\theta$ direction (we will follow this index convention throughout the rest of the paper).
Here $g$ with no subscripts refers to the metric on the 4-dimensional space-time, which obeys the Einstein equations (and $\sqrt{-g}$ is the square-root of its determinant). For generality, we have allowed the coupling constant $\alpha(\theta)$ for this Maxwell Lagrangian to be $\theta$-dependent.

As a side note, let us remark that there are some intriguing suggestions \cite{Lu_2302} that for a system with a global $\mathrm{U}(1)$ symmetry on a space-time $\partial M \times S^1$, with a 't Hooft anomaly described by \eqnref{eq:conservation_eqn} with $m\neq 0$, the condition $j^\theta = 0$ may in fact be enforced automatically once one applies the ``phase-space magnetic field'' \eqnref{eq:phase_space_magnetic_field}, so that the global symmetry gets upgraded to $\mathrm{LU}(1)$ automatically. Ref.~\cite{Lu_2302} only considered systems of non-interacting fermions, but if the result does hold more generally, it would suggest that in a holographic model we could just take the gauge field in the bulk theory to be a $\mathrm{U}(1)$ gauge field on $M \times S^1$, which would mean we could use the usual Maxwell action for a $\mathrm{U}(1)$ gauge field rather than \eqnref{eq:maxwell}. We leave exploration of this possibility for future work.

Next, it is also necessary to take into account the 't Hooft anomaly of $\mathrm{LU}(1)$. The standard way to implement a 't Hooft anomaly in the dual boundary theory is to include a Chern-Simons term for the bulk dynamical gauge field \cite{Witten_9802}. In particular, the anomaly equation \eqnref{eq:conservation_eqn} is obtained at the boundary of  the 5D Chern-Simons term
\begin{equation}
    \label{eq:CS}
    S_{CS} = \frac{m}{24\pi^2} \int_{M \times S^1} a \wedge da \wedge da.
\end{equation}
Note that, strictly speaking, the Chern-Simons term is not well-defined on a manifold with boundary, unless one imposes specific boundary conditions. However, the \emph{difference} in the action between two gauge-field configurations that have the same values on the boundary $\partial M \times S^1$ is well-defined, since this is equivalent to evaluating the Chern-Simons term on a closed manifold. For our purposes this will mostly be sufficient, but it will cause some difficulties in defining the relation between the bulk fields and the currents in the dual boundary theory, since according to the holographic dictionary, these are defined through variations of the bulk partition function with respect to the boundary values of the gauge field. We return to these issues in Section \ref{subsec:boundary}.

In summary, the dynamical gauge fields in the bulk are the metric $g$ and the $\mathrm{LU}(1)$ gauge field $a$, and the total action is given by
\begin{equation}
    \label{eq:total_action}
    S[g,a] = S_{\mathrm{CS}} + S_{\mathrm{Maxwell}} + S_{\mathrm{EH}},
\end{equation}
where the Chern-Simons action $S_{\mathrm{CS}}$ and the Maxwell action were defined above, and $S_{\mathrm{EH}}$ is the usual Einstein-Hilbert action for the metric:
\begin{equation}
    \label{eq:einstein_hilbert}
    S = \frac{1}{2\kappa^2} \int_{M} \sqrt{-g} \left(R + \frac{6}{L^2}\right) d^4 x,
\end{equation}
where $R$ is the Ricci scalar computed from the metric and $-6/L^2$ is the cosmological constant.
Note that since the Maxwell action does not depend on $a_\theta$, it is not possible to treat $a_\theta$ as a dynamical field in the bulk. Instead, we will just treat it as a fixed background.

    \subsection{Boundary conditions and identification of the currents in the dual QFT}
    \label{subsec:boundary}
To properly define the holographic correspondence, one needs to carefully consider the boundary conditions. Let us first observe that the classical equations of motion for the metric admit a solution which is asymptotically $\mathrm{AdS}_4$ near the boundary. We will adopt a coordinate system in which the asymptotic metric can be expressed as
\begin{equation}
    \label{eq:ads4}
    ds^2 = \frac{L^2}{r^2} (-dt^2 + dx^2 + dy^2 + dr^2),
\end{equation}
where the boundary is located at $r=0$.

Next we need to consider the asymptotic solutions for the $\mathrm{LU}(1)$ gauge field $a$ near $r=0$. Here our task is complicated by the presence of the Chern-Simons term in the action. For example, in the case of a $\mathrm{U}(1)$ gauge field in $\mathrm{AdS}_3$ with a Chern-Simons term $\sim \int a \wedge da$,  understanding the boundary conditions for holography becomes a somewhat involved topic,
see Ref.~\cite{Andrade_1111}. Fortunately, our task here is easier because in our case (unlike in the case of Maxwell-Chern-Simons in $\mathrm{AdS}_3$), one finds that the solutions have the same asymptotic scaling as $r \to 0$ with or without the Chern-Simons term, namely
\begin{equation}
\label{eq:a_asymptotic}
a_\mu = a^{(0)}_\mu + a^{(1)}_\mu r + \cdots,
\end{equation}
although the constraints on the coefficients $a^{(0)}$ and $a^{(1)}$ from the equations of motion may differ depending on the presence of the Chern-Simons term.
[To see that the solutions always have the asymptotic form \eqnref{eq:a_asymptotic}, just observe that with the metric \eqnref{eq:ads4}, the equations of motion do not have any singularity at $r=0$, hence the solutions must be analytic functions of $r$ at $r=0$.] This suggests that the holographic dictionary for a bulk Maxwell theory without a Chern-Simons term should simply carry over; that is, we should identify $a_\mu^{(0)}$ as the background gauge field applied in the dual boundary theory, while $a_\mu^{(1)}$ is the expectation value of the current operator in the boundary theory.

To make this argument more precise, first observe that in defining the action of the bulk theory, the asymptotic form \eqnref{eq:a_asymptotic} ensures that it will not be necessary to introduce any counterterms on the boundary to cancel divergent contributions at $r=0$, as is sometimes necessary in defining holographic duality.
However, another difficulty arises from the fact that to properly define the action, we need to define the Chern-Simons term in the presence of boundary, which has a certain ambiguity.

In this paper, we will seek to sidestep the issue in the following way. Suppose we consider \emph{two copies} of our system, with opposite sign of the anomaly coefficient $m$, and we impose that the background gauge field $A$ felt by the two copies should be the same. Then the combined system is dual to two copies of the gravitational theory, with opposite signs of the Chern-Simons level $m$ in \eqnref{eq:CS}, but with identical boundary values of the bulk gauge field $a$. Due to the different value of $m$, the bulk fields will evolve differently in the two copies. But the sum of the contributions to the action from the Chern-Simons terms of the two copies will not suffer from the ambiguity of a single copy, since evaluating this term is equivalent to evaluating the Chern-Simons action on a closed manifold obtained by gluing the two space-time manifolds together at their boundary. The doubled theory is only sensitive to responses of the original theory that are even under changing the sign of the anomaly coefficient $m$. Observe that in a microscopic lattice model of a metal, acting with a unitary particle-hole (i.e.\ ``charge conjugation'') operator on the microscopic Hamiltonian will lead to an opposite value of $m$ in the low-energy emergent theory without affecting the location of the Fermi surface. Therefore, we expect that any response that is even under such a particle-hole transformation, such as the linear electrical conductivity, will indeed be even under changing the sign of $m$. Responses that are odd under a particle-hole transformation, and hence under a change of sign of $m$, cannot be captured by the doubled theory, and would likely require more careful attention to the boundary conditions for the Chern-Simons term.

In any case, let us consider how to identify the currents of the dual boundary theory in the doubled system. First we observe that if we introduce the variation $\delta a$ of the gauge field, then by integrating by parts we see that the variation of the Maxwell term \eqnref{eq:maxwell} (in one of the copies) takes the form
\begin{equation}
    \label{eq:maxwell_variation}
      \int_{\partial M}d^3 x \int d\theta \sqrt{-g} \alpha(\theta)^{-1} A_\mu f^{r \mu} + \int_M d^4 x \int d\theta \sqrt{-g} \alpha(\theta)^{-1} A_\mu \partial_\mu f^{r \mu}.
\end{equation}
If we impose the classical equations of motion, then by definition the second term in \eqnref{eq:maxwell_variation} has to cancel the variation of the Chern-Simons term (one can verify that there is no boundary contribution coming from the Chern-Simons term in the doubled theory). Therefore, by taking the functional derivative with respect to $A_\mu$,  the current in the doubled theory is just given by sum of the contributions from the first term of \eqnref{eq:maxwell_variation} in the two copies, which gives: 
\begin{equation}
    j^\mu = -\frac{\delta S}{\delta A_\mu} = -\frac{\sqrt{-g}}{\alpha(\theta)} \left(f_{(1)}^{r \mu} + f_{(2)}^{r \mu}\right) \biggr|_{r=0}
\end{equation}
where the subscripts (1) and (2) refer to the fields in the two copies. This suggests that one should identify the current in the undoubled theory (modulo the caveats discussed above) as
\begin{equation}
    \label{eq:current_identification}
    j^\mu = -\frac{\sqrt{-g}}{\alpha(\theta)} f^{r \mu} \bigr|_{r=0}
\end{equation}
(which is the same as it would be in the absence of the Chern-Simons term).
Observe that the classical equations of motion in the bulk, \eqnref{eq:classical_eqs} imply that this current obeys the anomalous conservation equation \eqnref{eq:conservation_eqn}, with $j^\theta = 0$.

\subsection{The equilibrium solution in the bulk}
\label{subsec:equilibrium}
To describe the equilibrium properties of the system, we want to consider the dual QFT with global $\mathrm{LU}(1)$ symmetry at zero charge density (recall that if our theory represents the IR effective theory for some UV theory at nonzero charge density, this nonzero charge density is reflected in the emergent symmetry and anomaly of the IR theory, not its charge density).
Moreover, we will switch off all the background $\mathrm{LU}(1)$ gauge field, except that we still need to set $A_i(\theta) = k_i(\theta)$, where the spatial vector $\mathbf{k}(\theta)$ represents the Fermi surface momentum. In the gravitational theory this translates into the boundary condition for the bulk gauge field $a$. Recall that the necessity of including this ``phase space magnetic field'' was discussed in Section \ref{subsec:phase_space_magnetic_field}.

In this case, the solution of the classical equations of motion in the bulk are as follows. Firstly, the $\mathrm{AdS}_4$ metric \eqnref{eq:ads4} holds in the entire space-time, i.e.\ for all $r \geq 0$. Secondly, in the coordinate system in which the metric takes the form \eqnref{eq:ads4}, the $\mathrm{LU}(1)$ gauge field has components $a_x(\theta) = k_x(\theta), a_y(\theta) = k_y(\theta)$ (independently of $x$,$y$,$r$,and $t$), and the other components are zero. [Note that, while this gauge field has non-trivial gauge curvature $F_{\theta i}$, from the Maxwell action \eqnref{eq:maxwell} one sees that this component of the gauge curvature does not actually contribute to the stress tensor],  hence why the $\mathrm{AdS}_4$ metric remains a solution to Einstein's equations.)

The remainder of this paper will be devoted to computing responses of the dual QFT by considering perturbations to the equilibrium solutions. In order to make progress, we will only consider linear responses; this will allow us to linearize the equations of motion about the equilibrium solution.

\section{Results}
\label{sec:results}

\subsection{A preliminary remark: the UV cutoff scale}
\label{subsec:cutoff}
In this section we will present the results of solving the linearized classical equations of motion in the bulk. There is one point that needs to be kept in mind when interpreting these results, as follows. With respect to a physical lattice model of a metal, the model of an ersatz Fermi liquid that we have constructed is only supposed to be the effective IR theory. This places limitations on the regime in which the results we obtain will be meaningful. Specifically, we should focus on the response at frequency $\omega$, wavevectors $\mathbf{q}$, and temperature $T$, such that $|\omega|,|\mathbf{q}|,T$ are much smaller than some cut-off scale. 

As we will see, the solutions that we obtain appear to have a characteristic scale $u$, where
\begin{equation}
    u \sim \alpha(\theta) |m| |\partial_\theta \mathbf{k}(\theta)|.
\end{equation}
For example, if we assume an isotropic Fermi surface such that $\mathbf{k}(\theta) = k_F (\cos \theta, \sin \theta)$ and $\alpha(\theta) = \alpha$ is independent of $\theta$, then we have
\begin{equation}
    \label{eq:isotropic_u}
    u \sim |m| \alpha k_F.
\end{equation}
Thus, in this paper we will focus on the results in the regime $|\omega|,|\mathbf{q}|,T \ll u$. In other words, our goal will be to characterize the effective field theory that emerges in the deep IR at scales below $u$.

One could ask whether the results obtained in the holographic model are still meaningful for scales above $u$. We expect that the Fermi wavevector $k_F$ will place an upper bound on the scales for which the holographic model can be a useful description of the original microscopic lattice model. However, 
a condition for the electrodynamics of the bulk theory to be weakly coupled is [say in the isotropic case so that \eqnref{eq:isotropic_u} holds] that $\alpha \ll 1$. Therefore, if $m \sim 1$ then $u \ll k_F$. The holographic model could thus conceivably describe meaningful physics on scales greater than $u$. However, we will not focus on this regime in the current work.

\subsection{Charge responses at zero temperature}
\label{subsec:zeroT}
The linearized equations of motion for the $\mathrm{LU}(1)$ gauge field $a$ obtained from the action \eqnref{eq:total_action} do not contain any derivatives  with respect to $\theta$. Therefore, they can be solved independently at each $\theta$. Moreover, as the linearized equations of motion for the case of an AdS$_4$ metric turn out to be a system of ODEs with constant coefficients, they can be solved analytically in a straightforward way. However, as the form of the solution ends up being somewhat complicated in the general case, we will focus on the behavior for $|\omega|, |\mathbf{q}| \ll u$ as previously discussed in Section \ref{subsec:cutoff}. In that case, we show in Appendix \ref{appendix:solution} that one finds for the currents in the boundary theory in response to applied background gauge field\footnote{We have chosen the branch of the square root such that for real $\omega$ and $\mathbf{q}$, we have $\sqrt{-\omega^2 + q_\perp^2} = -i \operatorname{sgn}(\omega) \sqrt{\omega^2 - q_\perp^2}$ when $|\omega| > |q_\perp|$. while we just take the positive square root for $|\omega| < |q_\perp|$.}:

\begin{equation}
    \label{eq:jtjperp}
    \langle j^t \rangle = \langle j^\perp \rangle = \frac{m|\partial_\theta \mathbf{k}(\theta)|}{(2\pi)^2}\frac{i}{\omega - q_\perp} E_\perp  - i \alpha(\theta)^{-1} \frac{(\omega + q_\perp) q_\parallel }{(\omega - q_\perp) \sqrt{ -\omega^2 + q_\perp^2 }}(E_\parallel + B) + \cdots,
\end{equation}
\begin{equation}
    \label{eq:jpar}
    \langle j^\parallel \rangle = -i \alpha(\theta)^{-1} \frac{\omega+q_\perp}{\sqrt{-\omega^2+q_\perp^2}} (E_\parallel + B) + \cdots,
\end{equation}
where we defined the electric field $E_i = -i(q_i A_t + \omega A_i)$; and the magnetic field $B = i(q_x A_y - q_y A_x)$. Here we have written the spatial components of the vectors in terms of the components perpendicular to ($\perp$) and parallel to ($\parallel$) the Fermi surface: that is,
\begin{equation}
    \label{eq:some_stuff}
    q_\parallel = \frac{q_i w^i(\theta)}{\sqrt{w^j(\theta) w_j(\theta)}}, \quad \mathrm{etc.}
\end{equation}
where Roman letter indices such as $i$ take values in the two spatial dimensions, and we have defined $w^i(\theta) = \epsilon^{ij} \partial_\theta k_j(\theta)$ as before. To raise and lower spatial indices, we use the unit metric in the coordinate system $(x,y)$ in which the bulk metric takes the form \eqnref{eq:ads4}, i.e.\ the metric $ds^2 = dx^2 + dy^2$ (which is not the same as the bulk metric evaluated at the boundary, whose components diverge). This is also the metric that we use to evaluate $|\partial_\theta \mathbf{k}(\theta)| = w^i(\theta) w_i(\theta)$ in \eqnref{eq:jtjperp}.
In writing Eqs.~(\ref{eq:jtjperp}) and (\ref{eq:jpar}) we have assumed that the Chern-Simons level $m$ is positive; there are similar equations for $m < 0$ but with different signs.

Let us be more precise about what we mean by the ``$\cdots$'' in Eqs.~(\ref{eq:jtjperp}) and (\ref{eq:jpar}). One can argue  from the general structure of the equations of motion (see Appendix \ref{appendix:solution}) that the currents in linear response can be written as
\begin{equation}
    \label{eq:general_form}
    \langle j^\mu \rangle = \alpha(\theta)^{-1} \mathcal{J}^{\mu \nu} (\omega,\mathbf{q},u) A_\nu,
\end{equation}
where the function $\mathcal{J}$ depends \emph{only} on its explicit parameters $\omega$, $\mathbf{q}$ and $u$, and we have defined
\begin{equation}
    u = \frac{m \alpha(\theta) |\partial_\theta \mathbf{k}(\theta)|}{(2\pi)^2}
\end{equation}
Then we can expand $\mathcal{J}^{\mu \nu}$ in a power series in $1/u$:
\begin{equation}
    \label{eq:u_expansion}
    \mathcal{J}^{\mu \nu}(\omega,\mathbf{q},u) = \sum_{p=-1}^{\infty} \mathcal{J}^{\mu\nu}_{(p)}(\omega,\mathbf{q}) u^{-p}.
\end{equation}
Dropping the ``$\cdots$'' terms in Eqs.~(\ref{eq:jtjperp}) and (\ref{eq:jpar}) corresponds to keeping only the $p=0$ and $p=-1$ terms in this expansion.
The first term in \eqnref{eq:jpar} and the second term in \eqnref{eq:jtjperp} correspond to $p=0$ in \eqnref{eq:u_expansion}, while the first term in \eqnref{eq:jtjperp} corresponds to $p=-1$.

In the language of the renormalization group, $u$ serves as a UV cutoff for an effective field theory, and the $p>0$ terms will describe the effect of irrelevant operators, corresponding to the fact that they go to zero as $\omega/u$, $\mathbf{q}/u$ go to zero. The fact that there is a $p=-1$ term as well as a $p=0$ term is likely analogous to the following statement in Fermi liquid theory: when one defines the appropriate RG scaling, the effective theory contains a parameter $k_F / \Lambda$, where $\Lambda$ is the momentum cutoff scale, which flows to infinity under the RG flow. In this sense, Fermi liquid theory is not, strictly speaking, a fixed-point under RG [in which case one would have expected only the $p=0$ term to be present in the expansion \eqnref{eq:u_expansion}] but rather a one-parameter trajectory. As this behavior is tied to the fact that the low-energy excitations live on the Fermi surface rather than at zero momentum, one should expect a similar property to be true in our holographic model as well.

    If we keep only the leading-order term, i.e.\ the $p=-1$ term in \eqnref{eq:u_expansion}, in which case only the first term in \eqnref{eq:jtjperp} remains, then this exactly agrees with the result that would be obtained from the hydrodynamic equation of motion \eqnref{eq:hydrodynamic_eqn}, with the Fermi velocity $v_F$ equal to the speed of light $c$ in the bulk theory (set to 1 in our units). In particular, the pole at $\omega = q_\perp$ indicates a gapless propagating mode with velocity $v_F = 1$, but one which is chiral and directional since it can only move in one direction, perpendicular to the Fermi surface.
    In particular, as we noted in Section \ref{subsec:hydrodynamic}, this is the same result that would obtain in Fermi liquid theory, with the Landau interactions set to zero. [However, one should not view this result as suggesting that our theory is somehow ``weakly coupled'' like Fermi liquid theory, because as described in Section \ref{subsec:hydrodynamic}, the equation of motion \eqnref{eq:hydrodynamic_eqn} can be viewed as a general consequence of hydrodynamics, taking into account the conserved quantities associated with the $\mathrm{LU}(1)$ symmetry.]
    Meanwhile, the $p=0$ terms in the expansion have no analog in Fermi liquid theory and reflect non-Fermi liquid behavior.

Let us consider some particular limits of the general expressions Eqs.~(\ref{eq:jtjperp}) and (\ref{eq:jpar}). First of all, we compute the static susceptibility $\chi(\theta,\theta')$ for the $N(\theta)$ charges, which is defined by
\begin{equation}
    \chi(\theta,\theta') := \lim_{\mathbf{q} \to 0} \lim_{\omega \to 0} \frac{\delta \langle j^t(\theta') \rangle}{\delta A_t(\theta)} (\omega, \mathbf{q}).
\end{equation}
From \eqnref{eq:jtjperp} we find
\begin{equation}
    \label{eq:chi_theta}
\chi(\theta,\theta') = 
 \frac{m |\partial_\theta \mathbf{k}(\theta)|}{(2\pi)^2} \delta(\theta - \theta').
\end{equation}
In particular, we find that the total charge compressibility (i.e.\ the susceptibility of the total $\mathrm{U}(1)$ charge) is given by
\begin{equation}
    \chi = \iint \chi(\theta,\theta') d\theta d\theta' = \frac{m}{(2\pi)^2} \ell_F > 0,
\end{equation}
where $\ell_F = \int |\partial_\theta \mathbf{k}(\theta)| d\theta$ is the total length of the Fermi surface. The condition $\chi > 0$ is often used as a definition of ``compressibility''. In general this need not be equivalent to the definition of compressibility we gave in Section \ref{subsec:luttinger} and in the introduction, but in this model we find that the system is compressible in both senses.

 Another interesting case to look at is the regime of optical conductivity, where we set $B=0$ and then take the limit of $\mathbf{q} \to 0$ at fixed $\omega$. Then Eqs.~(\ref{eq:jtjperp}) and (\ref{eq:jpar}) (upon dropping the ``$\cdots$'') become
 \begin{equation}
    \label{eq:jtjperp_optical}
    \langle j^t \rangle = \langle j^\perp \rangle = \frac{m|\partial_\theta \mathbf{k}(\theta)|}{(2\pi)^2} \frac{i}{\omega} E_\perp,
 \end{equation}
 \begin{equation}
    \label{eq:jpar_optical}
    \langle j^\parallel \rangle =  \alpha(\theta)^{-1} E_\parallel.
 \end{equation}
Recall that these are the contributions to the currents from a particular point on the Fermi surface. To get the total current, we have to integrate over the whole Fermi surface; we assume that the electric and magnetic fields $\mathbf{E}$ and B are background gauge fields of the $\mathrm{U}(1)$ symmetry, which is to say that they are independent of $\theta$. One finds that the total charge density is zero, while the total current density is given by
\begin{equation}
    \langle j^i \rangle = \sigma^{ij}(\omega) E_j,
\end{equation}
with the conductivity tensor $\sigma(\omega)$ of the form
\begin{equation}
    \label{eq:sigma_omega}
    \sigma(\omega) = \mathcal{D} \frac{i}{\omega} + \sigma_{\mathrm{inc}},
\end{equation}
with the ``Drude weight''
\begin{equation}
\mathcal{D}^{ij} = \frac{m}{(2\pi)^2} \int \frac{w^i(\theta) w^j(\theta)}{|\mathbf{w}(\theta)|} d\theta,
\end{equation}
and
the frequency-independent ``incoherent conductivity''
\begin{equation}
    \sigma_{\mathrm{inc}}^{ij} = \int \frac{v^i(\theta) v^j(\theta)}{|\mathbf{w}(\theta)|^2} \alpha(\theta)^{-1} \, d\theta,
\end{equation}
where we defined $w^i(\theta) = \epsilon^{ij} \partial_\theta k_j(\theta)$ and $v_i(\theta) = \partial_\theta k_i(\theta)$, and we use the unit metric to raise and lower spatial indices as described below \eqnref{eq:some_stuff}.

Note that, as can be seen from Eqs.~(\ref{eq:jtjperp_optical}) and (\ref{eq:jpar_optical}), the two terms appearing in the conductivity \eqnref{eq:sigma_omega} have physically different origins -- the first term comes from the current that, at each point of the Fermi surface, flows perpendicular to the Fermi surface; while the second term comes from the current that flows \emph{parallel} to the Fermi surface. In a Fermi liquid the current only ever flows perpendicular to the Fermi surface, so this is another reflection of non-Fermi liquid behavior. Also, it is apparent from the solutions described in Appendix \ref{appendix:solution} that the first term, which is non-dissipative, arises from a bulk mode which decays exponentially with $r$ away from the boundary, while the second term, which is dissipative, arises from a mode which does not decay exponentially with $r$. This makes sense because in the bulk theory one can think of the energy lost in a dissipative process as falling into a black hole located (in the limit of zero temperature) at $r=\infty$, so a mode that decays exponentially with $r$ never reaches $r=\infty$ and hence will always be non-dissipative.

\subsection{Nonzero temperature}
To describe the model at nonzero temperature, we just need to replace the time direction of space-time by a compact Euclidean direction \cite{HQM}. As in Section \ref{subsec:equilibrium}, one finds that in the equilibrium state, the $\mathrm{LU}(1)$ gauge field does not enter into the equations of motion for the metric. As a result, the $\mathrm{AdS}_4$ metric \eqnref{eq:ads4} will simply be replaced by a thermal metric that has the same form as for a theory dual to a strongly-coupled (2+1)-D CFT, namely
\begin{equation}
    \label{eq:ads_thermal}
    ds^2 = \frac{L^2}{r^2} \left[ f(r) d\tau^2 + \frac{1}{f(r)} dr^2 + dx^2 + dy^2 \right],
\end{equation}
with 
\begin{equation}
    f(r) = 1 - \left( \frac{r}{r_+} \right)^3,
\end{equation}
and $r_+$ determined in terms of the temperature $T$ by
\begin{equation}
    r_+ = \frac{3}{4\pi} \frac{1}{T}.
\end{equation}
This reduces to asymptotic (Euclidean) $\mathrm{Ads}_4$ near the boundary, $r \to 0$, but the space-time ends at $r=r_+$, corresponding to a Euclidean version of a black hole event horizon.

The equations of motion for the $\mathrm{LU}(1)$ gauge field with the metric \eqnref{eq:ads_thermal} are no longer analytically solvable. However, we expect that the $p=-1$ term in the expansion \eqnref{eq:u_expansion} [that is, the Fermi-liquid-like term in \eqnref{eq:jtjperp}] will remain roughly unchanged for $T \ll u$. The reason is that this term arises from a mode that exponentially decays in the bulk for $r \gtrsim u$. Meanwhile, the thermal metric \eqnref{eq:ads_thermal} only differs appreciably from the Euclidean version of the zero-temperature metric when $r \gtrsim T^{-1}$. Therefore, if $T \ll u$ the mode should be unaffected by the nonzero temperature.

By contrast, the subleading contributions will likely be affected by nonzero temperature. Let us focus specifically on the optical conductivity. The ``Drude'' part of the optical conductivity, i.e.\ the first term in \eqnref{eq:sigma_omega}, should be unaffected for $T \ll u$ for the reasons described above. Meanwhile,
one can check that if one sets $\mathbf{q}=0$, then the $a_\parallel$ component of the gauge field decouples from $a_\perp$ and $a_t$, and obeys the same equation of motion as $\mathrm{U}(1)$ gauge field with a Maxwell action. Since it is the $a_\parallel$ component that is responsible for giving rise to the $\sigma_{\mathrm{incoherent}}$ term in \eqnref{eq:sigma_omega}, therefore this $\sigma_{\mathrm{incoherent}}$ will have the same dependence on $\omega$ and $T$ as in a holographic model of a (2+1)-D CFT at zero charge density, in which the bulk theory just has a $\mathrm{U}(1)$ gauge field with the Maxwell action and the metric \eqnref{eq:ads_thermal}. One can show \cite{Herzog_0701} that in fact, this always has the form
\begin{equation}
    \sigma_{\mathrm{incoherent}}(\omega,T) = \sigma_0,
\end{equation}
i.e.\ a constant independent of $\omega$ and $T$. This, however, is due to the special property of the self-duality of the Maxwell action and in general will not be the case if one introduces additional terms in the bulk action \cite{Myers_1010}. But more generally, the conductivity will obey the scale-invariance property of a quantum critical point in two spatial dimensions, i.e.
\begin{equation}
    \sigma_{\mathrm{incoherent}}(\omega,T) = f(\omega/T),
\end{equation}
for some scaling function $f$.

\section{Interpretation: what is the gravitational theory dual to?}
\label{sec:interpretation}
A difficulty with holographic models is that if, as we are doing here, one simply postulates an action for a bulk gravitational theory, it may be rather obscure what is the nature of the dual QFT. Nevertheless, in this instance we feel we are able to make a fairly good guess. The key observation is that, as we noted in Section \ref{subsec:equilibrium}, the metric takes the $\mathrm{AdS}_4$ form \eqnref{eq:ads4} throughout the entire bulk space-time, not just asymptotically near the boundary. This is the same form that one would expect for a quantum field theory that is dual to a strongly coupled (2+1)-D CFT (at zero charge density) in some large-$N$ limit. However, the charge response that we found in Section \ref{subsec:zeroT} does not take the form that one would expect in such a CFT. On the other hand, if we compute, for example, the entropy density as a function of temperature, then  the entropy density will be dominated by the gravitational contribution coming from the black hole in the metric \eqnref{eq:ads_thermal}, and therefore will have the same scaling with temperature as in such a CFT.

This motivates us to make the following proposal for the dual QFT, in the case where we set the Chern-Simons level $m$ (and hence, the anomaly coefficient of the dual QFT) equal to one: it corresponds to the IR effective theory resulting from coupling a spinless single-component Fermi liquid to a large-$N$ strongly coupled CFT. The fluctuations of the CFT will destroy the quasiparticles of the Fermi liquid, leading to a non-Fermi liquid, while still (one presumes) preserving the global $\mathrm{LU}(1)$ symmetry, at least in an emergent sense. Meanwhile, since $m \sim 1$ but $N \gg 1$, the Fermi liquid does not have enough degrees of freedom to significantly backreact on the CFT, corresponding to the statement in the dual theory that the bulk metric in equilibrium is unaffected by the $\mathrm{LU}(1)$ gauge field. Furthermore, the entropy density of the CFT will scale with some power of $N$, and therefore in the large-$N$ limit will dominate over any contribution from the Fermi surface.

The picture described above is also very reminiscent of the ``semi-holographic'' picture \cite{Faulkner_1001} that was developed in the context of some previous holographic models. In these models there is a small Fermi surface that does not satisfy Luttinger's theorem on its own. It was argued that the physics can be understood in terms of a Fermi liquid with the small Fermi surface coupled to a strongly coupled sector that contains most of the charge. By contrast, in the picture described above, the Fermi surface does contain all of the charge and the strongly coupled sector is at zero charge density.

One can compare this picture with other routes to obtaining non-Fermi liquids. For example, in Hertz-Millis type theories \cite{Hertz__1976,Millis__1993,Lee_1703}, one couples a Fermi liquid to a \emph{free} boson rather than a strongly coupled CFT; all the strong-coupling physics in such theories comes from the boson-fermion interactions.

Finally, let us also remark on the distinction with the SYK-inspired ``large-$N$ random-flavor'' models described in Ref.~\cite{Esterlis_1906,Aldape_2012,Esterlis_2103,Guo_2207,Shi_2208} \footnote{See also the previous work Ref.~\cite{Wang_1904}}, which are large-$N$ deformations of Hertz-Millis models. These seem to be natural candidates to have a holographic dual; for example, it has been argued that these models exhibit maximal quantum chaos in the large-$N$ limit \cite{Tikhanovskaya_2202}. However, the holographic model described in this paper cannot be dual to these theories. For one thing, in the random-flavor models one sends the number of fermion species [and hence, the anomaly coefficient $m$ for the $\mathrm{LU}(1)$ symmetry] to infinity. Meanwhile, in our holographic model we are free to just set $m=1$. Moreover, in the random-flavor models, in general the $\mathrm{LU}(1)$ charges will always have diverging susceptibilities in certain channels \cite{Shi_2204}, while in our holographic model the susceptibility remains finite, see \eqnref{eq:chi_theta}. Finally, we note that in these models one does not expect to have any current flowing in the direction parallel to the Fermi surface in the fixed-point theory \cite{Shi_2204}, in contrast to what we found in Section \ref{subsec:zeroT}. 

\section{Entanglement entropy and charge fluctuations}
\label{sec:entanglement}
A famous property of Fermi liquid theory \cite{Wolf_0503,Gioev_0504,Ding_1110,Calabrese_1111,Swingle_1112,Jiang_2209} in $d$ spatial dimensions is that the entanglement entropy in the ground state in a spatial region $M$ scales like $\sim L^{d-1} \log L$, where $L$ is a characteristic length scale of $M$; thus, the usual area law for entanglement entropy is violated logarithmically. One might ask whether our holographic model obeys the same property.

In holography, it is believed \cite{Faulkner_1307,Engelhardt_1408} that if the gravitational theory is sufficiently weakly coupled, such that one can ignore quantum fluctuations of the area,  the entanglement entropy of the dual QFT in a spatial region $M$  is given by
\begin{equation}
    \label{eq:entropy_formula}
    S(M) = \frac{2\pi}{\kappa} A(\mathcal{X}) + S_{\mathrm{ent}}(\mathcal{X}),
\end{equation}
where $\kappa$ is the gravitational constant appearing in the Einstein-Hilbert action \eqnref{eq:einstein_hilbert}; $\mathcal{X}$ is a codimension 1 surface in an equal-time slice of the bulk space-time, such the boundary of $\mathcal{X}$ coincides with the boundary of $M$; $A(\mathcal{X})$ is the area of $\mathcal{X}$ computed according to the metric of the bulk gravitational theory; and $S_{\mathrm{ent}}(\mathcal{X})$ is the entanglement entropy of the bulk quantum fields in the region delimited by the surface $\mathcal{X}$.
One is supposed to choose the \emph{extremal} surface, i.e.\ the surface which minimizes the right-hand side.

In order for the bulk theory to be weakly coupled, one is supposed to send $\kappa \to 0$. Therefore, in this limit, the first term of \eqnref{eq:entropy_formula} will dominate and one recovers the so-called ``Ryu-Takanagi'' formula \cite{Ryu_0603}. In this limit, the entanglement entropy is solely determined by the minimal area surfaces in the gravitational theory. Since in our model, with $d=2$, the metric takes the same form \eqnref{eq:ads4} as in theories dual to a (2+1)-D CFT, it follows that the contribution to the entanglement entropy coming from the first term of \eqnref{eq:entropy_formula} will obey the area law, $S(M) \sim L = L^{d-1}$.

However, it is still possible, and indeed we believe very likely, that there will be a $\sim L^{d-1} \log L$ contribution from the entanglement entropy coming from the second term in \eqnref{eq:entropy_formula} and in particular from the entanglement of the bulk $\mathrm{LU}(1)$ gauge field.  [Note that this would imply that $\kappa \to 0$ and $L \to \infty$ limits do not commute for the entanglement entropy], This is consistent with the picture of Section \ref{sec:interpretation}, in which one indeed expects the fermion contribution to the entanglement entropy to be subleading in $1/N$ compared to the contribution from the strongly coupled QFT. We will not attempt to compute this contribution to the  entanglement entropy in the current work. Instead, we will consider a related quantity, namely the charge fluctuations.

Let $Q_M$ be the operator that measures the total $\mathrm{U}(1)$ charge in the region $M$. Then we can consider the variance $(\Delta Q_M)^2 := \langle Q_M^2 \rangle - \langle Q_M \rangle^2$. In Fermi liquid theory, it turns out \cite{Gioev_0504,Swingle_1112} that $(\Delta Q_M)^2 \sim L^{d-1} \log L$.  This result tells us something about the correlations between $M$ and its complement, because at zero temperature the fluctuation of the \emph{total} charge of the ground state is zero, so $(\Delta Q_M)^2 > 0$ shows that the region $M$ and its complement must be correlated. Indeed, the fact that the charge fluctuations have the same scaling as the entanglement entropy suggests that the correlations between $M$ and its complement, which the entanglement entropy measures, are dominated by the charge fluctuations. Heuristically, one can view the fact that charge fluctuations grow faster than area law as related to the fact that (clean) Fermi liquids have zero DC resistivity in the limit of zero temperature, so it is very easy for the charge to ``slosh around'', as opposed to being bound locally in place as it would be in an insulator.

To compute the charge fluctuations in our holographic model, we can use the fluctuation-dissipation theorem to express the connected correlator $\langle n(\mathbf{q}) n(\mathbf{-q}) \rangle_c$ [or more generally, the $\theta$-resolved correlator $\langle n(\mathbf{q},\theta) n(\mathbf{-q},\theta') \rangle_c$] in terms of the retarded Green's function $G^R_{n(\mathbf{q},\theta) n(\mathbf{-q},\theta')} (\omega)$, which can be derived from the results in Section \ref{subsec:zeroT}. In the spirit of the renormalization group, the leading contribution to the equal-time correlator as $\mathbf{q} \to 0$ [and hence, the leading contribution to $(\Delta Q_M)^2$ as $L \to \infty$] should come from the most relevant operator. Therefore, we will keep only the $p=-1$ term in the expansion \eqnref{eq:u_expansion}. Observe that this term has exactly the same form as one would find in a non-interacting Fermi gas. Therefore, one expects to the get the same result for $(\Delta Q_M)^2$ as in a non-interacting Fermi gas.

In a non-interacting Fermi gas, it has been shown that the coefficient of $L^{d-1} \log L$ can be obtained exactly and has an elegant geometric expression  \cite{Gioev_0504,Calabrese_1111} . Suppose that $M$ is obtained by scaling a region $\Gamma \subseteq \mathbb{R}^d$ by a factor of $L$. Then one finds that
\begin{equation}
    \label{eq:fermi_fluctuations}
    (\Delta Q_M)^2 = \lambda_{\Gamma} L^{d-1} \log L + o(L^{d-1} \log L),
\end{equation}
with\footnote{Ref.~\cite{Gioev_0504} has an additional factor of $\log 2$ in this formula, but this is presumably an error; it does not appear in subsequent papers on the topic \cite{Calabrese_1111,Jiang_2209}.}
\begin{equation}
    \label{eq:lambdaGamma}
    \lambda_{\Gamma} =  \frac{m}{(2\pi)^{d+1}} \int_{\partial \Gamma} dA_x \int_{\mathcal{F}} dA_k |\mathbf{n}_x \cdot \mathbf{n}_{k}|.
\end{equation}
where $m$ is the multiplicity of the Fermi surface (i.e.\ the number of bands which have a Fermi surface at the same location), $\int_{\partial \Gamma} dA_x$ and $\int_{\mathcal{F}} dA_k$ denote surface integrals, $\mathcal{F}$ is the Fermi surface in momentum space, and $\mathbf{n}_x$ and $\mathbf{n}_k$ are the local unit normal vectors to the respective surfaces. We show in Appendix \ref{appendix:fluctuations} that Eqs.~(\ref{eq:fermi_fluctuations}) and (\ref{eq:lambdaGamma}) are indeed precisely what we get from the retarded Green's function computed in Section \ref{subsec:zeroT}, keeping only the $p=-1$ term in the expansion \eqnref{eq:u_expansion}.

In non-interacting Fermi gases, there are stronger results one can show regarding charge fluctuations. In particular \cite{Calabrese_1111}, all the higher cumulants of $Q_M$ fail to pick up any $\sim L^{d-1} \log L$ contribution and hence are suppressed relative to the variance $(\Delta Q_M)^2$ as $L \to \infty$. In other words, the charge fluctuations obey an approximately Gaussian distribution as $L \to \infty$.
It would be interesting to verify whether or not this holds in our holographic model. This would require computing nonlinear responses.

\section{Outlook}
\label{sec:outlook}
We do not want to claim that the particular model that we have studied here will itself explain everything about non-Fermi liquids. Nevertheless, it seems a much more viable starting point for studying non-Fermi liquids than previous holographic models, since it explicitly builds in the basic property of a Fermi surface satisfying Luttinger's theorem. An interesting future direction will be to consider adding perturbations to the strongly coupled quantum field theory that explicitly break the $\mathrm{LU}(1)$ symmetry, in order to model umklapp or disorder scattering; such perturbations have natural correspondences on the gravitational side through the holographic dictionary. One could also try to find perturbations that lead to an instability to a superconductor, or to another kind of ordered phase such as Ising-nematic.

One can also hope to use the model as a testing ground for hypothesized general statements about compressible metals; for example, according to the claims of Ref.~\cite{Else_2007}, if we explicitly break $\mathrm{LU}(1)$ but retain a $\mathbb{Z}^2 \times \mathrm{U}(1)$ subgroup corresponding to lattice translation symmetry and charge conservation, then the system should flow under RG to one in which the $\mathrm{LU}(1)$ symmetry is restored in an emergent sense, since compressible systems with lattice translation symmetry are supposed to have an infinite-dimensional emergent symmetry group. This should be a testable statement in our model.

Finally, the approach of designing holographic IR effective theories based on emergeability conditions or by targeting particular emergent symmetries and anomalies may be useful in other contexts beyond non-Fermi liquid metals. For example, a superfluid can be characterized \cite{Delacretaz_1908} by its emergent higher-form symmetry \cite{Gaiotto_1412}, which has a mixed anomaly with the 0-form charge $\mathrm{U}(1)$. Thus, one could hope to find a holographic model of a strongly coupled superfluid by studying an appropriate dynamical gauge field in the bulk with a Chern-Simons term. This idea was previously proposed as a future direction in Ref.~\cite{Delacretaz_1908}.

\section*{Acknowledgments}
I thank T. Senthil, Zhengyan Darius Shi, Meng Cheng, Subir Sachdev, Blaise Gout\'eraux, and Eric Mefford for helpful discussions. I was partly supported by the EPiQS initiative of the Gordon and Betty Moore foundation, grant nos.~GBMF8683 and GBMF8684. Research at Perimeter Institute is supported in part by the Government of Canada through the Department of Innovation, Science and Economic Development and by the Province of Ontario through the Ministry of Colleges and Universities.

\begin{appendices}
\section{Solving the linearized equations of motion}
\label{appendix:solution}
From the bulk action described in Section \ref{subsec:bulk_action}, we obtain the classical equations of motion for the gauge field in the bulk:
\begin{equation}
    \label{eq:classical_eqs}
    \partial_\mu [(\sqrt{-g}) f^{\mu \nu}] = \alpha(\theta)^{-1} \frac{m}{(2\pi)^2}\epsilon^{\nu \lambda \gamma \sigma} (\partial_\theta a_\lambda) (\partial_\gamma a_\sigma),
\end{equation}
where the indices range over the dimensions of space-time, but not the $\theta$ direction.
For the AdS$_4$ metric \eqnref{eq:ads4}, this conveniently reduces to the same equations of motion as in flat space (when expressed in terms of the covariant field-strength tensor $f_{\mu \nu}$) since the $(\sqrt{-g})$ factor in the left-hand side exactly cancels the components of the inverse metric that appear when we raise the indices of $f_{\mu \nu}$.

According to the discussion in Section \ref{subsec:equilibrium}, we introduce the equilibrium configuration of the gauge field,
\begin{equation}
    a_i^{(0)} = k_i(\theta),
\end{equation}
and then linearize \eqnref{eq:classical_eqs} in perturbations about this configuration. Furthermore, we take all fields to vary as $\sim e^{-i \omega t + k_i x^i}$ in the $t,x,y$ directions, and we choose a gauge in which we set $a_r = 0$. We obtain four equations of motion corresponding to setting $\nu = x,y,t$ or $r$ in \eqnref{eq:classical_eqs}. The first three can be collectively expressed as
\begin{align}
    \label{eq:Aeqn}
    \partial_r^2 \mathcal{A} + \mathcal{M} \partial_r \mathcal{A} + \Gamma \mathcal{A}  = 0,
\end{align}
where we defined
\begin{equation}
    \mathcal{A} = \begin{bmatrix} a_t \\ a_x \\ a_y \end{bmatrix}
\end{equation}
and
\begin{equation}
    \Gamma = \begin{bmatrix} -(q_x^2 + q_y^2) & -\omega q_x & -\omega q_y \\
    \omega q_x & \omega^2 - q_y^2 & q_x q_y \\ \omega q_y & q_x q_y & \omega^2 - q_x^2 \end{bmatrix}
\end{equation}
and
\begin{equation}
    \mathcal{M} = \begin{bmatrix} 0 & -u^x & -u^y \\ -u^x & 0 & 0 \\ -u^y & 0 & 0
    \end{bmatrix},
\end{equation}
with
\begin{equation}
u^i = \frac{m \alpha(\theta)}{(2\pi)^2} \epsilon^{ij} \frac{d}{d\theta} k_j(\theta).
\end{equation}
We will henceforth work in a coordinate system such that $u_x = u > 0, u_y = 0$.

    The fourth equation of motion can be written as
\begin{equation}
    \label{eq:the_fourth}
    i q_i F^{r i} - i\omega F^{r t} = -u^i e_i,
\end{equation}
where $e_i = -i \omega a_i - i q_i a_t$ are the components of the electric field in the $x$ and $y$ directions. Given the identifications \eqnref{eq:current_identification}, at $r=0$ this is precisely the statement of the anomalous conservation equation \eqnref{eq:conservation_eqn} in the dual boundary theory. Observe that if we take the derivative of \eqnref{eq:the_fourth} with respect to $r$, then it follows from the other three equations of motion. Therefore, the only effect of \eqnref{eq:the_fourth} will be to a fix a constant of integration. For the moment, therefore, we just consider the solutions of \eqnref{eq:Aeqn}.

Since this is a system of ODEs with constant coefficients, we can seek solutions of the form $\mathcal{A} \propto e^{\lambda r}$, which gives \begin{equation}
    (\lambda^2 + \lambda \mathcal{M} + \Gamma) \mathcal{A} = 0.
\end{equation}
This has a non-trivial solution for $\mathcal{A}$ when
\begin{equation}
\det(\lambda^2 \mathbb{I} + \lambda \mathcal{M} + \Gamma) = 0.
\end{equation}
Solving this equation gives a double root at $\lambda = 0$, and the other four solutions are
\begin{equation}
    \label{eq:lambda}
    \lambda = \sigma_1 \sqrt{-\omega^2 + q_x^2 + q_y^2 + \frac{u}{2} \left(u + \sigma_2 \sqrt{4 q_y^2 + u^2}\right)},
\end{equation}
where $\sigma_1$ and $\sigma_2$ can take the values $\pm 1$. If the argument of the outer square root is positive, then the boundary conditions at $r \to \infty$ require us to discard the solutions corresponding to $\sigma_1 = +1$ in \eqnref{eq:lambda}, since they blow up exponentially as $r \to \infty$, and retain only the exponentially decaying solutions corresponding to $\sigma_1 =-1$. If the argument of the outer square root is negative, then $\lambda$ is pure imaginary and the solutions correspond to radiative modes in the gravitational bulk that can propagate out to $r \to \infty$. In that case, the appropriate boundary condition to impose, consistent with causality, is that we keep only the mode that is radiating \emph{outwards} from $r=0$, where the external fields are applied, \emph{towards} $r = \infty$. This amounts to imposing that $\operatorname{sgn}(\operatorname{Im} \lambda) = \operatorname{sgn}(\omega)$.
To allow us to handle both cases at once, we will take the convention that when the argument of the square root is negative, we choose the branch such that $\sqrt{U} = -i \operatorname{sgn}(\omega)\sqrt{-U} $. Then we can always take the root with $\sigma_1 = -1$.

We remark that, since $\lambda=0$ is a double root, we have the solutions $\mathcal{A} = \mathcal{A}_0$ and $\mathcal{A} = \mathcal{A}_0 r + \mathcal{A}_1$,
where $\mathcal{A}_0$ and $\mathcal{A}_1$ satisfy $\Gamma \mathcal{A}_0 = 0$ and $\mathcal{M} \mathcal{A}_0 + \Gamma \mathcal{A}_1 = 0$. One can show that
\begin{equation}
    \mathcal{A}_0 = \begin{bmatrix} -\omega \\ q_x \\ q_y \end{bmatrix},
\end{equation}
and
\begin{equation}
    \mathcal{A}_1 = \frac{u}{\omega^2 - (q_x^2 + q_y^2)} \begin{bmatrix} q_x \\ -\omega \\ 0 \end{bmatrix}.
\end{equation}

Therefore, so far we have shown is that the general solution will take the form
\begin{equation}
    \label{eq:general_solution}
    \mathcal{A} = c_0 \mathcal{A}_0 + c_1 (\mathcal{A}_0 r + \mathcal{A}_1) + c_2 \mathcal{A}_2 e^{\lambda_2 r} + c_3 \mathcal{A}_3 e^{\lambda_3 r},
\end{equation}
for some integration constants $c_1,c_2,c_3,c_4$. Here, $\lambda$ and $\lambda'$ correspond to \eqnref{eq:lambda} upon setting $\sigma_1 = -1$ and [$\sigma_2 = 1$ (for $\lambda$) or $-1$ (for $\lambda'$)], and $\mathcal{A}_2$ and $\mathcal{A}_3$ are the corresponding eigenvectors.  Next we need to impose \eqnref{eq:the_fourth}. Because, as already mentioned, the $r$ derivative of \eqnref{eq:the_fourth} follows from the other three equations of motion, imposing \eqnref{eq:the_fourth} at one value of $r$ will be enough to imply that it is satisfied at all values of $r$. By sending $r \to \infty$, we find that we must set $c_1=0$.

The eigenvectors $\mathcal{A}_2$ and $\mathcal{A}_3$ have a somewhat complicated form, making the general calculation rather burdensome. However, a general statement that one can make is that the equations of motion only depend on $u$ and $(\omega,q_x,q_y)$. This justifies our statement that the  result for the currents will be of the form \eqnref{eq:general_form} [the factor of $\alpha(\theta)^{-1}$ comes from the final identification of the currents in the boundary theory, \eqnref{eq:current_identification}]. We ultimately relied on \textsc{Mathematica} to handle the tedious algebra, perform the expansion in $1/u$ described in Section \ref{subsec:zeroT}, and finally obtain the result given in Section \ref{subsec:zeroT} for the $p=-1$ and $p=0$ terms of the expansion \eqnref{eq:u_expansion} \footnote{The Mathematica notebook file used for the computations can be found at \url{https://arxiv.org/src/2307.02526/anc/odesoln.nb}.}. Here, however, in order to facilitate physical interpretation, we describe a simplified version of the calculation that can reproduce the leading-order terms in the result, i.e.\ the $p=-1$ term in \eqnref{eq:u_expansion}

As $u \to \infty$, to leading order \eqnref{eq:lambda} becomes
and
\begin{equation}
    \lambda = \pm u,
\end{equation}
\begin{equation}
    \lambda = \pm \sqrt{-\omega^2 + q_x^2}.
    \label{eq:some_lambda}
\end{equation}
One can show that to leading order, the corresponding eigenvectors take the form
\begin{equation}
    \begin{bmatrix} 1 \\ \pm 1 \\ 0
    \end{bmatrix}
\end{equation}
and
\begin{equation}
    \begin{bmatrix} 0 \\ 0 \\ 1 \end{bmatrix}
\end{equation}
respectively
[to this order, the eigenvectors corresponding to the pair of eigenvalues \eqnref{eq:some_lambda} with opposite signs are equal].
Thus, the $\lambda = \pm u$ modes are ``radial'' modes that involve the component of the gauge field perpendicular to the Fermi surface (i.e.\ in the coordinate system we are using, the $a_x$ component) as well as the time component, while the $\lambda = \pm \sqrt{-\omega^2 + q_x^2}$ modes are ``circumferential'' modes that involve the component of the gauge field parallel to the Fermi surface.

Thus, to leading order, the general solution \eqnref{eq:general_solution} (setting $c_1 = 0$) becomes
\begin{align}
    a_t &= c_2 e^{-u r} - c_0 \omega, \\
    a_x &= -c_2 e^{-u r} + c_0 q_x, \\
    a_y &= c_3 e^{-\sqrt{q_x^2 - \omega^2} r} + c_0 q_y.
\end{align}
We demand that at $r =0$, $a_t,a_x,a_y$ are equal to the applied background field $A_t,A_x,A_y$. This gives
\begin{equation}
    c_0 = \frac{A_t + A_x}{q_x - \omega}, \quad c_2 = \frac{q_x A_t + \omega a_x}{q_x - \omega}, \quad c_3 = A_y + \frac{(A_t + A_x) q_y}{\omega - q_x}.
\end{equation}
Finally, substituting into \eqnref{eq:current_identification} and keeping only the terms that are formally of order $p=-1$ in the expansion \eqnref{eq:u_expansion} gives the leading-order term in \eqnref{eq:jtjperp}.

\section{Computing charge fluctuations}
\label{appendix:fluctuations}
In this appendix we will derive the formulas Eqs.~(\ref{eq:fermi_fluctuations}) and (\ref{eq:lambdaGamma}) from the leading term in the retarded Green's function of the densities. From the results in Section \ref{subsec:zeroT}, keeping only the $p=-1$ term in the expansion \eqnref{eq:u_expansion}, we obtain
\begin{equation}
    \label{eq:GR}
    G^R_{n(\theta) n(\theta')} (\mathbf{q},\omega) = \frac{m}{(2\pi)^2}\frac{\mathbf{q} \cdot \mathbf{w}(\theta)}{\omega - q_\perp} \delta(\theta - \theta').
\end{equation}
This has the same form as a non-interacting Fermi gas in $d=2$ spatial dimensions, with the Fermi velocity equal to 1. For generality, let us consider general spatial dimension $d$ [in which case the Fermi surface is a $(d-1)$-dimensional manifold], and general Fermi velocity $v_F(\theta)$. Then the equivalent of \eqnref{eq:GR} is
\begin{equation}
    \label{eq:GR_general}
    G^R_{n(\theta) n(\theta')} (\mathbf{q},\omega) = \frac{m}{(2\pi)^d}\frac{\mathbf{q} \cdot \mathbf{w}(\theta)}{\omega - \mathbf{q} \cdot \mathbf{v}_F(\theta)} \delta^{d-1}(\theta - \theta'),
\end{equation}
where we defined $\mathbf{v}_F(\theta) = v_F(\theta) \mathbf{w}(\theta) / |\mathbf{w}(\theta)|$. In general dimension $\mathbf{w}(\theta)$ is defined according to
\begin{equation}
    w^i(\theta) = \epsilon^{ij_1 \cdot j_{d-1}} \partial_{\theta_1} k_{j_1}(\theta) \cdots \partial_{\theta_{d-1}} k_{j_d}(\theta),
\end{equation}
where $(\theta_1, \cdots, \theta_d)$ is some coordinate chart for the Fermi surface, and $\mathbf{k}(\theta)$ is the momentum of the Fermi surface as a function of $\theta$.

Now from the fluctuation-dissipation theorem, we have that the equal-time connected correlator of the densities is given by
\begin{equation}
    \langle n(\theta,\mathbf{q}) n(\theta',-\mathbf{q}') \rangle = \frac{1}{2\pi}\int_{-\infty}^{\infty} d\omega \, 2[1 + n_B(\omega)]\operatorname{Im} G^R_{n(\theta),n(\theta')}(\omega,\mathbf{q}) \times \delta^d(\mathbf{q} - \mathbf{q}').
\end{equation}
where the Bose factor $n_B(\omega)$ is defined by $n_B(\omega) := 1/(e^{\omega/T}-1)$. The only contribution to the imaginary part of \eqnref{eq:GR_general} comes from the pole at $\omega = \mathbf{q} \cdot \mathbf{v}_F(\theta)$ (which has to be resolved in the usual way by shifting $\omega$ infinitesimally off the real axis), so we obtain
\begin{equation}
    \operatorname{Im} G^R_{n(\theta),n(\theta')}(\omega,\mathbf{q}) = \frac{\pi m}{(2\pi)^d} \mathbf{q} \cdot \mathbf{w}(\theta) \, \delta[\omega - \mathbf{q} \cdot \mathbf{v}_F(\theta)] \, \delta^{d-1}(\theta - \theta').
\end{equation}
Hence, at zero temperature where $1+n_B(\omega)$ is just a Heaviside step function, we find
\begin{equation}
    \label{eq:nq_correlator}
    \langle n(\theta,\mathbf{q}) n(\theta',-\mathbf{q}') \rangle = \frac{m}{(2\pi)^d} \mathbf{q} \cdot \mathbf{w}(\theta) \Theta(q_\perp) \times \delta^{d-1}(\theta - \theta') \, \delta^d(\mathbf{q} - \mathbf{q}'),
\end{equation}
where $\Theta$ is the Heaviside step function, and as before $q_\perp$ is the component of $\mathbf{q}$ parallel to $\mathbf{w}(\theta)$ (i.e.\ perpendicular to the Fermi surface). To avoid UV divergences, we will introduce a UV cutoff by multiplying the right-hand side of \eqnref{eq:nq_correlator} by an additional factor of $e^{-a q_\perp}$, which defines the cutoff scale $a$. Then, taking the Fourier transform gives
\begin{equation}
    \label{eq:nx_correlator}
    \langle n(\theta,\mathbf{x}) n(\theta',\mathbf{x}')\rangle_c = \frac{m|\mathbf{w}(\theta)|}{(2\pi)^{d+1}} \frac{1}{(x_\perp - x'_\perp + ia)^2} \delta^{d-1}(\mathbf{x}_\parallel - \mathbf{x}_\parallel') \delta^{d-1}(\theta - \theta').
\end{equation}
where $x_\perp$ is the component of $\mathbf{x}$ parallel to $\mathbf{w}(\theta)$, and $\mathbf{x}_{\parallel}$ is the projection of $\mathbf{x}$ into the plane parallel to the Fermi surface, i.e.\  normal to $\mathbf{w}(\theta)$.

Finally, we can compute the charge fluctuation in a region $M$:
\begin{equation}
    (\Delta Q_M)^2 = \int_M d^d \mathbf{x} \int_M d^d \mathbf{x}' \int d^{d-1}\theta \int d^{d-1}\theta' \, \langle n(\theta,\mathbf{x}) n(\theta',\mathbf{x}') \rangle_c.
\end{equation}
Substituting \eqnref{eq:nx_correlator}, we find
\begin{multline}
    (\Delta Q_M)^2 = \frac{m}{(2\pi)^{d+1}}\int d^{d-1} \theta |\mathbf{w}(\theta)| \int_{M_{d-1}} d^{d-1} \mathbf{x}_{\parallel} \int_{x_\perp^{-}(\mathbf{x}_{\parallel})_M}^{x_\perp^{+}(\mathbf{x}_{\parallel})_M} dx_\perp \int_{x_\perp^{-}(\mathbf{x}_{\parallel})_M}^{x_\perp^{+}(\mathbf{x}_{\parallel})_M} dx_{\perp}' \\ \times \frac{1}{(x_\perp - x_\perp' + ia)^2},
\end{multline}
where $[x_\perp^{-}(\mathbf{x}_{\parallel})_M, x_\perp^{+}(\mathbf{x}_{\parallel})_M]$ denotes the intersection of $M$ with the 1-dimensional line of fixed $\mathbf{x}_{\parallel}$ (here for simplicity we have assumed that the Fermi surface is convex so that this intersection is just a single interval, but this is not essential), and we only integrate $\mathbf{x}_{\parallel}$ over the region $M_{d-1} \subseteq \mathbb{R}^{d-1}$ such that this intersection is non-empty. Performing the integral over $dx_\perp$ and $dx_\perp'$ gives
\begin{equation}
\log(x_\perp^{+} - x_\perp^{-} + ia) + \log(x_{\perp}^{-} - x_\perp^{+} + ia) - 2 \log(ia).
\end{equation}
Up to subleading contributions this is just $2\log[ (x_\perp^{+} - x_\perp^{-})/a ]$. 
Hence, we find
\begin{equation}
    \label{eq:DeltaQM_1}
    (\Delta Q_M)^2 = \frac{2m}{(2\pi)^{d+1}}\int d^{d-1}\theta |\mathbf{w}(\theta)| \int_{M_{d-1}} d^{d-1} \mathbf{x}_{\parallel} \log\left( \frac{\Delta x_\perp(\mathbf{x}_\parallel)_M}{a} \right).
\end{equation}
 Now suppose that our region $M$ is obtained from a region $\Gamma$ by rescaling by a factor $L$. Then we can write \eqnref{eq:DeltaQM_1} as
\begin{equation}
    (\Delta Q_M)^2 = \frac{2m}{(2\pi)^{d+1}} L^{d-1} \int d^{d-1}\theta |\mathbf{w}(\theta)| \int_{\Gamma_{d-1}} d^{d-1} \mathbf{x}_{\parallel}\left[ \log L + \log\left( \frac{\Delta x_\perp(\mathbf{x}_\parallel)_\Gamma}{a} \right) \right].
\end{equation}
Hence we find that
\begin{equation}
    (\Delta Q_M)^2 = \lambda_\Gamma L^{d-1} \log L + o(L^{d-1} \log L),
\end{equation}
with the coefficient
\begin{equation}
    \lambda_\Gamma = \frac{2m}{(2\pi)^{d+1}} \int d^{d-1}\theta \int_{\Gamma_{d-1}} d^{d-1} \mathbf{x}_{\parallel} |\mathbf{w}(\theta)|.
\end{equation}
We can recognize this as an equivalent way of writing \eqnref{eq:lambdaGamma}.

\end{appendices}

\printbibliography

\end{document}